\documentclass[aps,preprint,showpacs,preprintnumbers,amsmath,amssymb]{revtex4}
\usepackage{graphicx}

\begin{document} 

\def\beq{\begin{equation}}
\def\eeq{\end{equation}}
\def\beqn{\begin{eqnarray}}
\def\eeqn{\end{eqnarray}}
\def\btimes {\mbox{\boldmath $\times$}}
\def\bbox {\mbox{\boldmath $\box$}}
\def\bvarphi {\mbox{\boldmath $\varphi$}}
\def\bPhi {\mbox{\boldmath $\Phi$}}
\def\bPsi {\mbox{\boldmath $\Psi$}}
\def\ed{\end{document}}

\def\veps {{\varepsilon}}
\def\I {{\bf I}}
\def\II {{\bf II}}
\def\III {{\bf III}}
\def\IV {{\bf IV}}
\def\V {{\bf V}}
\def\VI {{\bf VI}}
\def\J {{\bf J}}
\def\K {{\bf K}}
\def\H {{\bf H}}
\def\R {{\bf R}}
\def\S {{\bf S}}
\def\O {{\bf 1}}

\def\E {{\bf E}}
\def\1 {{\bf 1}}
\def\2 {{\bf 2}}
\def\3 {{\bf 3}}
\def\P {{\bf P}}
\def\r {{\bf r}}
\def\x {{\bf x}}
\def\y {{\bf y}}

\def\k {{\bf k}}
\def\p {{\bf p}}
\def\n {{\bf n}}
\def\A {{\bf A}}
\def\bv {{\bf v}} 
\def\AAN {$\!\!\!$ A$^{^{\!\!\!\!\! {\tiny {\circ}}}}$}
\def\aaN {$\!\!$ a$^{^{\!\!\!\! {\tiny {\circ}}}}$}

\title{Cold atoms in real-space optical lattices} 

\author{Ofir E. Alon\footnote{E-mail: ofir@pci.uni-heidelberg.de}, Alexej I. Streltsov and Lorenz S. Cederbaum}
\affiliation{Theoretische Chemie, Physikalisch-Chemisches Institut, Universit\"at Heidelberg,\\
Im Neuenheimer Feld 229, D-69120 Heidelberg, Germany}

\begin{abstract}
Cold atoms in optical lattices are described in {\it real space}
by multi-orbital mean-field Ans\"atze.
In this work we consider four typical systems:
(i) spinless identical bosons, (ii) spinor identical bosons (iii),
Bose-Bose mixtures, and (iv) Bose-Fermi mixtures 
and derive in each case the corresponding multi-orbital mean-field 
energy-functional and working equations.
The notions of {\it dressed} Wannier functions and Wannier spinors 
are introduced and the equations defining them are presented and discussed.
The dressed Wannier functions are the set of orthogonal, translationally-equivalent  
orbitals which minimizes the energy of the Hamiltonian including 
boson-boson (particle-particle) interactions.
Illustrative examples of dressed Wannier functions
are provided for spinless bosonic atoms and mixtures in one-dimensional optical lattices.
\end{abstract}

\maketitle
\section{Introduction}

The first experimental realization of the superfluid to Mott-insulator quantum phase transition 
in cold bosonic atoms trapped by a three-dimensional (3D) optical lattice \cite{IB1_nature},
following the theoretical suggestion in \cite{Jaksch1_PRL},
has developed into one of the most active research subjects in cold-atom physics nowadays, 
see the recent reviews \cite{Markus_rev,Lewenstein_rev} and references therein.
Other experimental studies soon followed, 
including, e.g., demonstration of the superfluid to Mott-insulator transition 
in effective one-dimensional (1D) optical lattices \cite{TE1_PRL} 
and observation of the so-called Tonks-Girardeau gas 
thus entering deep into the strongly-interacting regime \cite{IB2_nature}.
Very recently, loading of atomic Bose-Fermi mixtures into 
3D optical lattices has been achieved \cite{Essliner2,Bongs}
where coherence properties have been measured and studied.

Trapping of cold atoms in optical lattices
(as well as in other trap geometries) 
occurs in {\it real-space}.
We therefore wish to attack the properties of cold atoms in optical lattices in real space.
Recently, a multi-orbital best-mean-field ansatz for spinless identical bosonic systems
has been derived \cite{LA_OAL_PLA}.
It allowed us to obtain {\it quantitative} value of the standard superfluid to Mott-insulator
transition of weakly-interacting spinless bosons in deep 1D optical lattices  
and led us to predict a wealth of quantum phases and excitations of strongly-interacting bosons 
in optical lattices \cite{Zoo}, 
and novel phenomena associated with fragmentation and fermionization of bosons in traps \cite{LA_ALN_PRA,Pathway}.
Very recently, in a first application of the multi-orbital best-mean-field approach
to bosonic mixtures, we have described demixing scenarios of bosonic mixtures in optical lattices 
from macroscopic to microscopic length scales \cite{BB_archive}.
The anticipation that the multi-orbital best-mean-field approach would 
continue to serve as a valuable tool for cold-atom systems in optical lattices 
and other traps serves as the main motivation of the present work.
To be more specific, the purposes of this paper are:
(i) to extend the multi-orbital mean-field approach originally developed for spinless identical bosons
 to spinor condensates and to Bose-Fermi atomic mixtures, 
  and to explicitly derive the formalism for Bose-Bose atomic mixtures;
(ii) to analyze the multi-orbital approach in the context of optical lattices, 
     highlighting the role and effects of particle-particle interactions and
     translational symmetry on the properties of the solutions (orbitals).

The structure of the paper is as follows.
In section II we present the 'original' multi-orbital mean-field for spinless bosons
and discuss some of its properties in the context of bosons in optical lattices.
Attention is given to the notion of dressed Wannier functions
which is introduced and demonstrated with the aid of illustrative 
numerical examples in 1D.
In section III we extend the idea of a multi-orbital mean-field for bosons
to spinor condensates. Working equations are explicitly 
derived and discussed in the context of optical lattices.
In section IV we proliferate the idea of a multi-orbital mean-field for single-species bosons
to bosonic mixtures. Working equations are explicitly derived.
Attention is given to the dressed Wannier functions which arise due
to intra- and inter-species interactions.
Illustrative numerical examples in 1D are provided.
In section V we treat Bose-Fermi mixtures,
and explicitly derive the corresponding working equations
of the multi-orbital mean-field approach and discuss them in the context
of trapped Bose-Fermi mixtures in optical lattices.
Finally, in section VI we present concluding remarks and some outlook.

\section{Multi-orbital mean-field for spinless identical bosons\\ in optical lattices} 

\subsection{Theory}

Our starting point is the many-body Hamiltonian describing $N$ spinless identical bosons 
of a single species in an optical lattice (trap),
\beq\label{ham}
 \hat H(\r_1,\r_2,\ldots,\r_{N}) =  \sum_{i=1}^{N}
 \left[ h(\r_i) +
 \sum_{j>i}^{N} \lambda_0V(\r_i-\r_j) \right].
\eeq
Here, $h(\r)$ is the one-body Hamiltonian,
containing kinetic and potential energy (the optical lattice) terms, 
$\r_i$ is the coordinate of the $i$-th particle,
and $\lambda_0V(\r_i-\r_j)$ 
describes the pairwise interaction between the $i$-th and $j$-th bosonic atoms
where $\lambda_0$ is a parameter which measures the strength of the interparticle interaction.
Another assumption, pertinent to physically realistic potentials,
is that two-body matrix elements associated with $V(\r_i-\r_j)$ (see below) are finite numbers. 
As is well known from electron-structure theory,
even for potentials as singular and long-ranged as the Coulomb potential
these matrix elements are finite quantities.

As mentioned above, we are going to obtain a {\it real-space} mean-field picture
of the quantum state of cold atoms in the optical lattice.
How are we going to achieve that?
To this end, we attach an {\it orbital} to each of the $N$ atoms.
The simplest choice is the Gross-Pitaevskii approach, 
for which all bosons reside in the same orbital,
\beq\label{GP}
 \Psi(\r_1,\r_2,\ldots,\r_N) = \phi(\r_1) \phi_2(\r_2) \cdots \phi(\r_N).
\eeq
There are, however, many other situations for bosons \cite{LA_OAL_PLA}.
Generally, we may take $n_1$ bosons to reside in one orbital, $\phi_1(\r)$,
$n_2$ bosons to reside in a second orbital, $\phi_2(\r)$, and so on,
distributing the $N$ atoms among $n_{orb}>1$ orthonormal orbitals.
At the other end to the Gross-Pitaevskii approach lies the situation where each
boson in the optical lattice resides in a different orbital, i.e., $n_{orb}=N$ \cite{Zoo,Pathway}.
More formally, the multi-orbital mean-field wavefunction for $N$ interacting spinless bosons is 
given by the following {\it single-configuration} wavefunction \cite{LA_OAL_PLA},
\beq\label{general_MF}
  \Psi(\r_1,\r_2,\ldots,\r_N) = \hat{\cal S} 
 \left\{ \phi_1(\r_1) \phi_2(\r_2) \cdots \phi_N(\r_N) \right\}, 
\eeq 
where $\hat{\cal S}$ is the symmetrization operator.
Note that the Gross-Pitaevskii equation is a specific case of Eq.~(\ref{general_MF}) 
when all orbitals are alike, namely $n_{orb}=1$.

To proceed for any set of $n_{orb}$ orbitals and a corresponding set of occupations $\{n_i\}$,
the expectation value of the Hamiltonian (\ref{ham}) with the
multi-orbital mean-field wavefunction (\ref{general_MF}) readily reads \cite{LA_OAL_PLA}, 
\beqn\label{BMF_functional}
   E = & & \sum_i^{n_{orb}} n_i \Bigg[ \int \phi_i^\ast(\r) h(\r) \phi_i(\r) d\r -
 \lambda_0 \frac{n_i+1}{2} 
 \int \!\! \int \phi_i^\ast(\r) \phi_i^\ast(\r') V(\r-\r')
\phi_i(\r) \phi_i(\r') d\r d\r' + \nonumber \\
  + & & \frac{1}{2}\sum_{j}^{n_{orb}} \lambda_0 n_j 
 \int \!\! \int \phi_i^\ast(\r) \phi_j^\ast(\r') V(\r-\r') \left\{1+{\mathcal P}_{\r\r'}\right\}
 \phi_i(\r) \phi_j(\r') d\r d\r' \Bigg], \
\eeqn
where ${\mathcal P}_{\r\r'}$ permutes the $\r$ and $\r'$ coordinates of two 
bosons (particles) appearing to the right of it.
Note the plus sign preceding the operator ${\mathcal P}_{\r\r'}$
which is due to the (bosonic) symmetrization operator.
Also, recall here our working assumption that the two-body 
matrix elements are finite quantities.

The ground state is obtained by minimizing the energy-functional $E$
with respect to its arguments which are the {\it number} of different orbitals $n_{orb}$,
the set of occupations $\{n_i\}$ and, of course,
the shape of the orbitals $\phi_i(\r)$.
This results in a set of $n_{orb}$ coupled equations that
have to be solved self-consistently \cite{LA_OAL_PLA},
\beqn\label{BMF_equations_general}
 & & \left\{ h(\r) - \lambda_0(n_i+1) J_i(\r) +
\sum_{j}^{n_{orb}} \lambda_0 n_j \left[ J_{j}(\r) + K_{j}(\r) \right] \right\} \phi_i(\r) =
 \sum_j^{n_{orb}} \mu_{ij} \phi_j(\r), \nonumber \\
 & & \ \ i=1,\ldots,n_{orb},
\eeqn
where the direct and exchange potentials are given by
\beqn\label{BMF_1_body}
& &  J_j(\r) = \int \phi_j^\ast(\r') V(\r-\r')
 \phi_j(\r') d\r', \nonumber \\
& &  K_j(\r) = \int \phi_j^\ast(\r') V(\r-\r') {\mathcal P}_{\r\r'} \phi_j(\r') d\r'. \
\eeqn
The Lagrange multipliers $\mu_{ij}$ appearing on the r.h.s. of Eq.~(\ref{BMF_equations_general}) 
are introduced in order to ensure orthonormality of the orbitals, 
$\int \phi^\ast_i(\r)\phi_j(\r) d\r = \delta_{ij}$, 
and satisfy the relations $n_i \mu_{ij}=n_j \mu_{ji}$. 
It is instructive to mention that, unlike the case of Hartree-Fock
equations for fermions \cite{Szabo_book}, 
the off-diagonal Lagrange multipliers cannot be removed in general; 
Also see subsequent sections
and in particular section V dealing with 
Bose-Fermi mixtures.

Let us analyze some properties of the energy-functional (\ref{BMF_functional}) 
as far as they are needed here,
assuming $V(\r-\r')$ to be a repulsive physical interaction.
For more details see \cite{LA_OAL_PLA,Pathway}.
For weakly-interacting atoms, the energy-functional $E$ is minimized
when all bosons reside in the same orbital (the Gross-Pitaevskii orbital),
namely for $n_{orb}=1$.
When $\lambda_0$ is increased,
the two-body, mean-field
energy terms start to dominate $E$ and the occupations $\{n_i\}$ 
of the bosons would like to become smaller in order to
minimize the energy. This leads to the occupation of more
and more orbitals in the ground state.
Eventually, the number of orbitals becomes equal to the number of bosons, namely all $n_i=1$.
In other words, the energy functional (\ref{BMF_functional}) is minimized 
by a set of $n_{orb}=N$ orthonormal orbitals, 
a result that holds in general trap potentials and for any physical 
repulsive interaction \cite{LA_OAL_PLA,Pathway}.

In what follows we would like to study in some detail the nature and appearance of the
multi-orbital mean-field solutions in optical lattices.
In particular we are interested in how the interaction affects, or dresses the interaction-free orbitals.
In optical lattices,
symmetry sets in and 
it is anticipated that Eq.~(\ref{BMF_equations_general}) possesses solutions
which reflect the translational and other possible symmetries of the lattice.
As usual, periodic boundary conditions are assumed, 
thus we employ a super-cell of a large 
number $N_w$ of potential wells (unit cells).
To be more specific, 
we consider an optical lattice with a commensurate filling factor of  
$f=\frac{N}{N_w}$ bosons per site, 
and examine solutions to Eq.~(\ref{BMF_equations_general})
described by $n_{orb}=N_w$ orbitals, i.e., $n_i=f \ \forall i$.
In this case and when exploiting the translational symmetry of the lattice, 
we are dealing with a set of $N_w$ orthonormal orbitals
which may be generated by duplicating an orbital $\phi(\r)$ located in, say, the zeroth unit cell, 
and translating its replica to all other unit cells.  
Denoting by $\R_i$ the set of lattice vectors which generate the entire
lattice from the zeroth unit cell (set $\R_1=0$), 
the following relations then hold between the orbitals:
$\phi(\r-\R_i)=\phi_i(\r) \ \forall i$.
With these conventions, Eq.~(\ref{BMF_equations_general}) boils down to, $i=1,\ldots,N_w$: 
\beq\label{BMF_dressed_Wanneir}
 \left\{ h(\r) - (f+1) \lambda_0 J(\r) +
\sum_j^{N_w} \lambda_0 f \left[ J(\r-\R_j) + K(\r-\R_j) \right] \right\} 
\phi(\r) = \sum_j^{N_w} \mu_{ij} \phi(\r+\R_i-\R_j).
\eeq

What are the solutions of Eq.~(\ref{BMF_dressed_Wanneir})?
If the inter-particle interaction vanishes, i.e., $\lambda_0=0$, 
then Eq.~(\ref{BMF_dressed_Wanneir}) reduces to, $i=1,\ldots,N_w$:  
\beq\label{BMF_dressed_Wanneir_0}
h(\r) \phi(\r) = \sum_j^{N_w} \mu_{ij} \phi(\r+\R_i-\R_j),
\eeq
which is nothing but the text-book definition of Wannier functions \cite{vol1}.
From this, we conclude that the solutions of Eq.~(\ref{BMF_dressed_Wanneir}) 
may be considered as boson-dressed Wannier functions,
namely, Wannier functions that are 
dressed by the interaction between the bosons.
From a complementary perspective,
the dressed Wannier functions are the set of orthogonal, translationally-equivalent  
orbitals which minimizes the energy of the Hamiltonian including 
boson-boson interaction.

\subsection{Illustrative examples of boson-dressed Wannier functions}

As mentioned above,
the application of the 'original' multi-orbital mean-field to
spinless bosons in optical lattices has already demonstrated 
fascinating results, see \cite{Zoo}. 
These include {\it quantitative} determination of the superfluid
to Mott-insulator transition in deep 1D optical lattices 
and a zoo of Mott-insulator phases and excitations involving higher bands.
Here, as illustrative numerical examples, 
we would like to concentrate on
the concept of boson-dressed Wannier functions,
which appear, e.g., in the calculation of {\it self-consistent}
Mott-insulator phases \cite{Zoo}, 
and study how interaction dresses the bear
Wannier functions in different scenarios.

We concentrate on 1D optical lattices,
$u(x)=u_0 \sin^2(kx)$, where $k$ is the wave vector.
Optical lattice depths, $u_0$, are expressed 
in terms of the recoil energy, $E_R=\hbar^2 k^2/2m$,
where $m$ is the mass of the atoms.
The usual contact interaction is employed,
$\lambda_0V(x-x')=\lambda_0\delta(x-x')$,
where $\lambda_0$ is related to the scattering 
length and the transverse harmonic confinement \cite{Maxim_PRL}. 
The strength of the inter-particle interaction is expressed via 
the dimensionless parameter $\gamma=m \lambda_0/\hbar^2 \bar n$
(the ratio of interaction and kinetic energies),
where $\bar n$ is the linear density.
We denote the (commensurate) filling factor by $f$.

Let us begin by dressing the Wannier functions in a deep optical lattice, $u_0=25E_R$.
For a commensurate filling of $f=1$, $N_w$ orbitals, 
and a weak interaction strength of $\gamma=0.00776002$,
the dressed Wannier functions are shown in Fig.~1a.
For the above parameters as can be seen in the figure, 
the spatial overlap between two neighboring functions is small.
For commensurate filling of $f=2$, $N_w$ orbitals,  
and a much stronger interaction strength $\gamma=12.7418$,
the dressed Wannier functions are shown in Fig.~1b.
Since there are now two interacting bosons per orbital,
the dressed Wannier functions are wider and lower.
In addition, two neighboring functions possess a larger spatial overlap,
compare Figs.~1a and 1b, 
in spite of the lattice being deep.

Next, we consider the dressed Wannier functions in a shallow optical lattice, $u_0=E_R$.
For a commensurate filling of $f=1$, $N_w$ orbitals, 
and an interaction strength of $\gamma=3.491$,
the dressed Wannier functions are shown in Fig.~2.
For shallow lattices the (dressed) Wannier functions extend beyond the 
next-nearest neighbor sites and 
the spatial overlap between two neighboring functions becomes substantial, 
compare Figs. 1 and 2.
Consequently, dressing due to particle-particle interaction 
in shallow optical lattices, even for filling factors as low as $f=1$, 
is expected to have a larger impact than in deep optical lattices.

So far, we dressed the Wannier functions of the lowest band in an optical lattice 
(technically, by taking $N_w$ orbitals).
We would like before concluding this section 
to present an example of two-band bosons-dressed Wannier functions.
To this end, 
we consider an optical lattice with a commensurate filling of $f=2$,
$2N_w$ orbitals, and an interaction strength of $\gamma=12.7609$.
Minimizing the multi-orbital energy-functional we obtain a set of
$2N_w$ {\it equivalent} Wannier functions which are connected one
to the other by translation and/or reflection symmetry, see Fig.~3.
Of course, all $2N_w$ dressed Wannier functions are orthogonal one to the others.
Since the bosons repel each other quite strongly, 
the dressed Wannier functions (orbitals)
would like to reduce their spatial overlap in order to minimize
the interaction in the multi-orbital energy-functional (\ref{BMF_functional}).
These $2N_w$ dressed Wannier functions are substantially different
from the bear Wannier functions of the first and second band.
We remind that, in the absence of particle-particle interaction, 
the bear Wannier functions in a deep 1D optical lattice 
resemble the two lowest eigenstates of the Harmonic-oscillator Hamiltonian
which clearly overlap in each unit cell.

\section{Multi-orbital mean-field for spinor identical bosons} 

The above theory of multi-orbital mean-field for spinless identical bosons can
be readily expanded to treat spinor condensates.
This task is carried out below for spin-1 (spinor) identical bosons.
Our starting point is the following many-body Hamiltonian describing $N$ spin-1 bosons 
of a single species in a trap (optical lattice), see, e.g., \cite{Ho,Pu_1999,Pethick_book,Wadati},
\beq\label{ham_spin}
 \hat \H(\r_1,\r_2,\ldots,\r_{N}) =  \sum_{i=1}^{N}
 \left\{ h(\r_i) \O^{(i)} + \sum_{j>i}^{N} V(\r_i-\r_j) 
 \left[\lambda_0 \O^{(i)}\otimes \O^{(j)} + \lambda_2 \sum_{\nu=x,y,z} 
\S_\nu^{(i)} \otimes \S_\nu^{(j)} \right]\right\}.
\eeq
Here, $h(\r)$ is the one-body Hamiltonian which is assumed to be the same for
all three components, $\O^{(i)}$ is a three-by-three unit matrix operating 
in hyperfine space on the $i$-th boson. 
$V(\r_i-\r_j)$ describes the angular-momentum-preserving interparticle interaction between the 
$i$-th and $j$-th atoms (for simplicity it is assumed to be a local interaction),
where $\lambda_0$ and $\lambda_2$ measure the 
strength of the interparticle interaction in
the spin-independent (symmetric) and spin-dependent 
(asymmetric) parts of (\ref{ham_spin}).
We remark that the contact inter-particle interaction 
$V(\r_i-\r_j)=\delta(\r_i-\r_j)$ is usually taken in the 
literature \cite{Ho,Pu_1999,Pethick_book,Wadati}.
For ease of presentation and to avoid displaying multiple tensor-product terms, 
$\O^{(k)}$ terms with $k\ne i,j$ are not shown in the Hamiltonian (\ref{ham_spin}).
Finally, the matrices $\S_\nu$ are the usual angular-momentum-$1$ matrices ($\hbar=1$),
\beq\label{AM1}
\S_x=\frac{1}{\sqrt{2}}\begin{pmatrix}0 & 1 & 0 \cr 1 & 0 & 1 \cr 0 & 1 & 0 \cr\end{pmatrix}, \ \
\S_y=\frac{1}{\sqrt{2}}\begin{pmatrix}0 & -i & 0 \cr i & 0 & -i \cr 0 & i & 0 \cr\end{pmatrix}, \ \
\S_z=\frac{1}{\sqrt{2}}\begin{pmatrix}1 & 0 & 0 \cr 0 & 0 & 0 \cr 0 & 0 & -1 \cr\end{pmatrix}. \ \
\eeq

In the same manner that we treated spinless bosons,
we are going to describe the state of the spin-1 condensate in real space. 
To this end, we again attach a one-body function to each of the $N$ atoms.
But this time, for spin-1 bosons, it is a vector of orbitals, or a spinor with three components:
$\bPhi(\r)=\begin{pmatrix}\phi_+(\r) \cr \phi_0(\r) \cr \phi_-(\r)\end{pmatrix}$.
The normalization condition reads $\int \! \bPhi^\dag(\r) \cdot \bPhi(\r) d\r = 1$. 
The simplest choice is the Gross-Pitaevskii approach,
for which all spin-1 bosons reside in the same spinor,
\beq\label{GP_spin}
\bPsi(\r_1,\r_2,\ldots,\r_N) = \bPhi(\r_1) \otimes \bPhi(\r_2) \otimes \cdots \otimes \bPhi(\r_N).
\eeq
This is the generally employed mean-field ansatz for spin-1 condensates \cite{Ho,Pu_1999,Pethick_book,Wadati}.
Moreover, in certain cases (e.g., $\lambda_2 \ll \lambda_0$) 
the so-called single-mode approximation 
has been invoked in which the spatial dependence of each component is alike, 
namely $\bPhi(\r)$ is taken as $\bPhi(\r)=\phi(\r)\begin{pmatrix}a_+ \cr a_0 \cr a_-\end{pmatrix}$,
where $a_\pm,a_0$ are scalars satisfying the normalization condition $|a_+|^2+|a_0|^2+|a_-|^2=1$.
 
The most general mean-field is obtained by placing $n_1$ spin-1 
bosons in one spinor, $\bPhi_1(\r)$,
$n_2$ spin-1 bosons in a second spinor, $\bPhi_2(\r)$, and so on,
distributing the $N$ spin-1 atoms among $n_{orb}>1$ orthonormal spinors.
Note that the individual components of different spinors need not be orthogonal to one another.
This defines the multi-orbital mean-field ansatz for spinor condensates,
\beq\label{general_MF_spin}
  \bPsi(\r_1,\r_2,\ldots,\r_N) = \hat{\cal S} 
 \left\{ \bPhi_1(\r_1) \otimes \bPhi_2(\r_2) \otimes \cdots \otimes \bPhi_N(\r_N) \right\}. 
\eeq  
As in the spinless case discussed before,
the corresponding Gross-Pitaevskii ansatz (\ref{GP_spin}) is a specific case of 
Eq.~(\ref{general_MF_spin}) when all spinors are alike.

To proceed, let us derive the multi-spinor energy-functional.
For any set of $n_{orb}$ spinors and a corresponding set of occupations $\{n_i\}$,
the expectation value of the Hamiltonian (\ref{ham_spin}) with the
multi-orbital (multi-spinor) mean-field wavefunction (\ref{general_MF_spin}) reads
{\small{
\beqn\label{BMF_functional_spin}
& &  E_{\mathit{spin-1}} = \sum_i^{n_{orb}} n_i 
\Bigg\{ \int \bPhi_i^\dag(\r) \cdot h(\r)\bPhi_i(\r) d\r - 
 \frac{n_i+1}{2} \int \!\! \int d\r d\r' V(\r-\r') \times \nonumber \\
& & \bigg[ \lambda_0 \left(\bPhi_i^\dag(\r) \cdot \bPhi_i(\r)\right) 
\left(\bPhi_i^\dag(\r') \cdot \bPhi_i(\r')\right) 
 + \lambda_2 \sum_\nu \left(\bPhi_i^\dag(\r) \cdot \S_\nu \bPhi_i(\r)\right)
 \left(\bPhi_i^\dag(\r') \cdot \S_\nu \bPhi_i(\r')\right) \bigg] +  \frac{1}{2}\sum_{j}^{n_{orb}} n_j \times
 \nonumber \\
  & & \int \!\! \int d\r d\r' V(\r-\r') 
 \bigg\{ \lambda_0  \left[ \left(\bPhi_i^\dag(\r) \cdot \bPhi_i(\r)\right) 
\left(\bPhi_j^\dag(\r') \cdot \bPhi_j(\r')\right) + 
\left(\bPhi_i^\dag(\r) \cdot \bPhi_j(\r)\right) 
\left(\bPhi_j^\dag(\r') \cdot \bPhi_i(\r')\right) \right] + \nonumber \\
& & \lambda_2 \sum_\nu \left[\left(\bPhi_i^\dag(\r) \cdot \S_\nu \bPhi_i(\r)\right)
 \left(\bPhi_j^\dag(\r') \cdot \S_\nu \bPhi_j(\r')\right) +
 \left(\bPhi_i^\dag(\r) \cdot \S_\nu \bPhi_j(\r)\right)
 \left(\bPhi_j^\dag(\r') \cdot \S_\nu \bPhi_i(\r')\right)\right]\bigg\}\Bigg\}. \
\eeqn
}}

Once the multi-orbital energy-functional is formulated, 
the ground state is obtained by minimizing $E_{\mathit{spin-1}}$ 
with respect to the {\it number} of different spinors $n_{orb}$,
the set of occupations $\{n_i\}$ and, of course,
the shape of the spinors $\bPhi_i(\r)$.
This results in a set of $n_{orb}$ coupled equations for the spinors $\bPhi_i(\r)$ 
[or, $3n_{orb}$ coupled equations for their respective 
components $\phi_{+,i}(\r)$, $\phi_{0,i}(\r)$ and $\phi_{-,i}(\r)$]
that have to be solved self-consistently and
bear a similar appearance to spinless-boson case, 
see Eq.~(\ref{BMF_equations_general}),
\beqn\label{BMF_equations_general_spin}
 & & \left\{ h(\r)\O - (n_i+1) \J_i(\r) +
\sum_{j}^{n_{orb}} n_j \left[ \J_{j}(\r) + \K_{j}(\r) \right] \right\} \bPhi_i(\r) =
 \sum_j^{n_{orb}} \mu_{ij} \bPhi_j(\r), \nonumber \\
 & & \ \ i=1,\ldots,n_{orb}.
\eeqn
Here, the direct and exchange matrix-potentials are defined by,
{\small{
\beqn\label{BMF_1_body_spin}
& &  \J_j(\r) = \int V(\r-\r') d\r'
\left[\lambda_0\left(\bPhi_j^\dag(\r') \cdot \bPhi_j(\r')\right)\O +
   \lambda_2\sum_\nu \left(\bPhi_j^\dag(\r') \cdot \S_\nu \bPhi_j(\r')\right) \S_\nu \right], \\
& &  \K_j(\r)\bPhi_i(\r) = 
\int V(\r-\r') d\r'
\left[\lambda_0\left(\bPhi_j^\dag(\r') \cdot \bPhi_i(\r')\right)\O +
  \lambda_2\sum_\nu \left(\bPhi_j^\dag(\r') 
\cdot \S_\nu \bPhi_i(\r')\right) \S_\nu \right]\bPhi_j(\r). \nonumber \
\eeqn
}}
The Lagrange multipliers $\mu_{ij}$ appearing on the r.h.s. of Eq.~(\ref{BMF_equations_general_spin}) 
are introduced in order to ensure orthonormality of the spinors, 
$\int \bPhi^\ast_i(\r) \cdot \bPhi_j(\r) d\r = \delta_{ij}$, 
and satisfy the relations $n_i \mu_{ij}=n_j \mu_{ji}$. 
We remind that the individual components of different spinors need not be orthogonal to one another.
As for the spinless case, the off-diagonal Lagrange multipliers cannot be removed in general. 

Analysis of the spin-1 multi-orbital energy-functional $E_{\mathit{spin-1}}$ can be drawn along
similar lines to that performed in the 'original', spinless multi-orbital case.
For instance, we may enquire what is the ``pathway to fermionization'' of 
a spinor condensate when $\lambda_0$ and $\lambda_2$ are enlarged
by increasing the transverse confinement in an effective 1D trap. 
Clearly, from some interaction strength on the energy-functional 
(\ref{BMF_functional_spin}) is minimized when every spin-1 atom
resides in a different spinor. 
Beyond that, detailed investigations are needed to determine how the
individual components $\phi_{\pm,i}(\r)$ and $\phi_{0,i}(\r)$ behave in the strongly-interacting limit.

In the present context of optical lattices and analogously to the spinless case, 
dressed {\it Wannier spinors}, $\bPhi(\r)$, are obtained due to particle-particle interaction.
By considering an optical lattice with a commensurate filling factor of  
$f=\frac{N}{N_w}$ spin-1 bosons per site, 
the equation defining the dressed Wannier spinors is obtained 
from Eq.~(\ref{BMF_equations_general_spin}) 
following the same lines employed in section II. 
The result resembles Eq.~(\ref{BMF_dressed_Wanneir}) 
where spinor quantities substitute there 
spinless ones, $i=1,\ldots,N_w$: 
\beq\label{BMF_dressed_Wanneir_spin}
 \left\{ h(\r) \O - (f+1) \J(\r) +
\sum_j^{N_w} f \left[ \J(\r-\R_j) + \K(\r-\R_j) \right] \right\} 
\bPhi(\r) = \sum_j^{N_w} \mu_{ij} \bPhi(\r+\R_i-\R_j).
\eeq
Note that the coupling between the different components 
$\phi_\pm(\r),\phi_0(\r)$ arising for $\lambda_2 \ne 0$
results in {\it each} component being dressed differently.
This finding suggests that distinct dressed Wannier functions
may become important for describing the different components 
of spinor condensates in optical lattices.

\section{Multi-orbital mean-field for Bose-Bose mixtures} 

So far, we treated bosonic systems made of one type of atoms.
The 'original' multi-orbital mean-field in the single-species case has 
been quite successful for bosons in optical lattices and other trap potentials.
Very recently, in a first application of the multi-orbital best-mean-field approach
to bosonic mixtures, we have described demixing scenarios of bosonic mixtures in optical lattices 
from macroscopic to microscopic length scales \cite{BB_archive}.
Accordingly, the purpose of this section is to provide an explicit and 
detailed derivation of the Bose-Bose multi-orbital formalism 
(i.e., for a mixture of two kinds of spinless bosons) 
and to discuss some of its general 
properties in the context of bosonic mixtures 
in optical lattices.

\subsection{Theory}

Our starting point is the many-body Hamiltonian describing $N=N_A+N_B$ spinless bosons,
$N_A$ bosons of type $A$ and $N_B$ bosons of type $B$, in a trap (optical lattice), 
\beq\label{ham_BB}
\hat H_{AB}(\r_1,\ldots,\r_{N_A},\r'_1,\ldots,\r'_{N_B}) =  \hat H_A + \hat H_B +
 \sum_{i}^{N_A}\sum_{p}^{N_B}\lambda_{AB}V_{AB}(\r_i-\r'_p).
\eeq
Here, $\hat H_A$ is the single-species Hamiltonian (\ref{ham})
where the corresponding quantities are denoted there by the subscript $A$,
e.g., $\lambda_AV_A(\r_i-\r_j)$ denotes the intra-species interaction.
Similarly, $\hat H_B$ is the single-species Hamiltonian of the $B$-type bosons
where the corresponding quantities are denoted there by the subscript $B$.
Finally, $\lambda_{AB}V_{AB}(\r_i-\r'_p)$
denotes the inter-species interaction between the $i$-th $A$ boson and the $p$-th
$B$ boson, where $\lambda_{AB}$ measures the strength of the inter-species interaction.

As mentioned above, we are going to obtain a {\it real-space} mean-field picture
of the quantum state of the {\it bosonic mixture} in the trap.
Similarly to the single-species case,
we attach an {\it orbital} to each of the $N=N_A+N_B$ atoms.
The simplest choice is governed by the so-called two-component Gross-Pitaevskii approach, 
for which all $A$-type bosons reside in one orbital
and all $B$-type bosons reside in another orbital, see, e.g., \cite{Pethick_book,BB_Ho,BB_Esry,BB_Pu},
\beq\label{GP_BB}
 \Psi(\r_1,\ldots,\r_{N_A},\r'_1,\ldots,\r'_{N_B}) 
 = \phi(\r_1) \phi(\r_2) \cdots \phi(\r_{N_A}) \times \psi(\r'_1)\psi(\r'_2) \cdots \psi(\r'_{N_B}).
\eeq
There can, however, 
be many other situations for the bosonic mixture,
by writing a multi-orbital mean-field 
ansatz for {\it each} one of the species in the mixture \cite{BB_archive}. 
Generally, we may take $n_1$ $A$-type bosons to reside in one orbital, $\phi_1(\r)$,
$n_2$ $A$-type bosons to reside in a second orbital, $\phi_2(\r)$, and so on,
distributing the $N_A$ atoms among $n_{orb}>1$ orthonormal orbitals.
Independently, we may take $m_1$ $B$-type bosons to reside in one orbital, $\psi_1(\r)$,
$m_2$ $B$-type bosons to reside in a second orbital, $\psi_2(\r)$, and so on,
distributing the $N_B$ atoms among $m_{orb}>1$ orthonormal orbitals.
We point out that due to the distinguishability of
the $A$- and $B$-type bosons, no relations are assumed between the $\phi$'s
and $\psi$'s orbitals which, therefore, are {\it a priori} allowed to overlap.
All together, the multi-orbital mean-field wavefunction for the $N=N_A+N_B$
bosons in the mixture is given by the multiplication of two single configurations \cite{BB_archive}:
\beq\label{general_MF_BB}
 \Psi(\r_1,\ldots,\r_{N_A},\r'_1,\ldots,\r'_{N_B}) =
 \hat{\cal S} \left\{ \phi_1(\r_1)\phi_2(\r_2) \cdots \phi_{N_A}(\r_{N_A}) \right\} \times
 \hat{\cal S} \left\{ \psi_1(\r'_1)\psi_2(\r'_2) \cdots \psi_{N_B}(\r'_{N_B}) \right\}.
\eeq 
Note that the two-component Gross-Pitaevskii approach \cite{Pethick_book,BB_Ho,BB_Esry,BB_Pu}
is a specific case of
Eq.~(\ref{general_MF_BB}) when all $A$-type orbitals are alike
and all $B$-type orbitals are alike, namely $n_{orb}=m_{orb}=1$.

To proceed for any set of $n_{orb}$ and set of $m_{orb}$
orbitals and corresponding sets of occupations $\{n_i\}$ and $\{m_i\}$,
the expectation value of the Hamiltonian (\ref{ham_BB}) with the
multi-orbital mean-field wavefunction (\ref{general_MF_BB}) readily reads:
\beq\label{BMF_functional_BB}
E_{AB} = E_A + E_B + \sum_i^{n_{orb}} \sum_p^{m_{orb}} \lambda_{AB} n_i m_p
  \int \!\! \int \phi_i^\ast(\r) \psi_p^\ast(\r') V_{AB}(\r-\r') 
                 \phi_i(\r) \psi_p(\r') d\r d\r'. 
\eeq
Here, $E_A$ is the single-species multi-orbital energy-functional,
see Eq.~(\ref{BMF_functional}), where the $A$ boson quantities are inserted therein,
and similarly $E_B$ is the single-species energy-functional of the $B$-type bosons.
The last term in the Bose-Bose energy-functional $E_{AB}$ represents
the interaction between the two species.

The ground state is obtained by minimizing the energy-functional $ E_{AB}$
with respect to the numbers of different $A$- and $B$-type orbitals $n_{orb}$ and $m_{orb}$,
the sets of occupations $\{n_i\}$ and $\{m_p\}$ and, of course,
the shape of the orbitals $\phi_i(\r)$ and $\psi_p(\r)$.
This results in a set of $n_{orb}+m_{orb}$ coupled equations that
have to be solved self-consistently,
\beqn\label{BMF_equations_general_BB}
 & & \left\{ h_A(\r) - \lambda_A(n_i+1) J_{A,i}(\r) +
\sum_{j}^{n_{orb}}\lambda_A n_j \left[J_{A,j}(\r) + K_{A,j}(\r)\right] 
+ \sum_{q}^{m_{orb}} \lambda_{AB} m_q J_{AB,q}(\r) \right\} \phi_i(\r) = \nonumber \\
& & \qquad = \sum_j^{n_{orb}} \mu_{A,ij} \phi_j(\r), \qquad \qquad 
 i=1,\ldots,n_{orb}, \nonumber \\
 & & \left\{ h_B(\r) - \lambda_B(m_p+1) J_{B,p}(\r) +
\sum_{q}^{m_{orb}}\lambda_B m_q \left[J_{B,q}(\r) + K_{B,q}(\r)\right] 
+ \sum_{j}^{n_{orb}} \lambda_{AB} n_j J_{BA,j}(\r) \right\} \psi_p(\r) = \nonumber \\
 & & \qquad = \sum_q^{m_{orb}} \mu_{B,pq} \psi_q(\r), \qquad \qquad 
 p=1,\ldots,m_{orb}. \
\eeqn
Here, $J_{A,j}(\r)$ and $K_{A,j}(\r)$ are the intra-species direct and exchange potentials,
see Eq.~(\ref{BMF_1_body}), where the $A$ boson quantities are inserted therein,
and similarly $J_{B,q}(\r)$ and $K_{B,q}(\r)$ are defined for the $B$-type bosons.
The inter-species direct-like potentials are given by
\beqn\label{BMF_1_body_BB}
& &  J_{AB,q}(\r) = \int \psi_q^\ast(\r') V_{AB}(\r-\r')
 \psi_q(\r') d\r', \nonumber \\
& &  J_{BA,j}(\r) = \int \phi_j^\ast(\r') V_{AB}(\r-\r') 
 \phi_j(\r') d\r'. \
\eeqn
The Lagrange multipliers $\mu_{A,ij}$ are introduced in order to ensure orthonormality
of the $A$-type orbitals, $\int \phi^\ast_i(\r)\phi_j(\r) d\r = \delta_{ij}$, 
and satisfy the relations $n_i \mu_{A,ij}=n_j \mu_{A,ji}$.
Analogously and {\it independently}, 
the Lagrange multipliers $\mu_{B,pq}$ are 
introduced to ensure orthogonality of the $B$ orbitals,
$\int \psi^\ast_p(\r)\psi_q(\r) d\r = \delta_{pq}$,
and satisfy the relations $m_p \mu_{B,pq}=m_q \mu_{B,qp}$.
As for the single-species case, 
the off-diagonal Lagrange multipliers 
of {\it both} sets cannot be removed in general.

Having at hand the energy-functional $E_{AB}$ we can study the changes in the ground-state
as the inter-species and/or intra-species interactions are varied.
Very recently, as mentioned above, in a first application of the 
Bose-Bose energy-functional (\ref{BMF_functional_BB}) 
and as a specific case-study of  the ``pathway to fermionization'' of bosonic mixtures,
we have described demixing scenarios of bosonic mixtures in optical lattices \cite{BB_archive}.
For completeness, we briefly discuss below how the physics of 
Ref.~\cite{BB_archive} emerges from the Bose-Bose energy-functional (\ref{BMF_functional_BB}). 
The main advantageous of the multi-orbital mean-field approach is 
that the orbitals $\{\phi_i(\r)\}$ and $\{\psi_p(\r)\}$
are determined self-consistently and thus depend on
the intra- and inter-species interactions and on the
density of each bosonic species, 
whereas in previous treatment of bosonic mixtures in optical lattices,
see, e.g., Ref.~\cite{Altman} and references therein, orbitals
(Wannier functions) whose shapes do not 
depend on these parameters were employed.

Consider a given bosonic mixture with some fixed intra-species interaction
strengths $g_A$, $g_B$ and inter-species repulsion $g_{AB}$.
We can minimize the energy-functional (\ref{BMF_functional_BB}) and find the ground state,
i.e., the set of orbitals  $\{\phi_i(\r)\}$, $\{\psi_p(\r)\}$
and their respective occupations $\{n_i\}$, $\{m_p\}$ that minimize $E_{AB}$.
For $g_{AB} \ll g_A,g_B$ the single-species terms $E_A$ and $E_B$
dominate $E_{AB}$ because the inter-species interaction term
[the third term on the r.h.s. of Eq.~(\ref{BMF_functional_BB})] is much smaller.
Consequently, both species can completely mix and spread all over the optical lattice.
Moreover, the physics of each species is determined by its own parameters almost independently
of the other species state.
Increasing $g_{AB}$ influences the physics of {\it both} species.
In order to reduce the inter-species interaction energy in (\ref{BMF_functional_BB}),
the orbitals $\{\phi_i(\r)\}$ and $\{\psi_p(\r)\}$ reduce their overlap in space.
When $g_{AB} \gg g_A,g_B$ we expect this overlap to vanish, which means
that the two species completely separate and occupy different regions in space.
In between these two extreme cases, 
there are many intriguing possibilities where the {\it self-consistent} 
orbitals of the $A$- and $B$-type bosons 
'intermingle' in between one another, 
what results in demixing scenarios 
on various length scales \cite{BB_archive}.

Next, let us discuss the dressing of Wannier functions in 
a bosonic mixture trapped in an optical lattice.
To this end, we consider an optical lattice with commensurate filling factors of $f_A$ and $f_B$, 
namely $n_i=f_A \ \forall i$ and $m_p=f_B \ \forall p$.
In this case, by exploiting the translational symmetry of the lattice we arrive at the final result
for the equations defining the dressed Wannier functions 
$\phi(\r)$ and $\psi(\r)$ of the bosonic mixture which read, $i=1,\ldots,N_w$ and $p=1,\ldots,N_w$:
\beqn\label{BMF_dressed_Wanneir_BB}
 & & \Bigg\{ h_A(\r) - \lambda_A(f_A+1) J_A(\r) +
\sum_{j}^{N_w} \Big[\lambda_A f_A \left[J_A(\r-\R_j) + K_A(\r-\R_j) \right] + \nonumber \\ 
 & & \qquad + \lambda_{AB} f_B J_{AB}(\r-\R_j)\Big] \Bigg\} \phi(\r) 
 = \sum_j^{N_w} \mu_{A,ij} \phi(\r+\R_i-\R_j), \nonumber \\
 & & \Bigg\{ h_B(\r) - \lambda_B(f_B+1) J_B(\r) +
\sum_{q}^{N_w} \Big[\lambda_B f_B \left[J_B(\r-\R_q) + K_B(\r-\R_q) \right] + \nonumber \\ 
 & & \qquad + \lambda_{AB} f_A J_{BA}(\r-\R_q)\Big] \Bigg\} \psi(\r)
 = \sum_q^{N_w} \mu_{B,pq} \psi(\r+\R_p-\R_q). \
\eeqn
Obviously, the inter- and intra-species interactions dress together
the $A$- and $B$-type-boson Wannier functions.
In the absence of inter-species interaction,
Eq.~(\ref{BMF_dressed_Wanneir_BB})
boils down to two independent equations for the single-species
dressed Wannier functions of the $A$ bosons and of the $B$ bosons, 
see Eq.~(\ref{BMF_dressed_Wanneir}),
whereas in the absence of all interactions
it boils down to two independent equations for the
text-book Wannier function, Eq.~(\ref{BMF_dressed_Wanneir_0}).

\subsection{Illustrative examples of Bose-Bose dressed Wannier functions}

As mentioned above,
the application of the Bose-Bose multi-orbital energy-functional (\ref{BMF_functional_BB})
to bosonic mixtures in optical lattices has already demonstrated 
fascinating results in connection with demixing, see \cite{BB_archive}.
Here, as illustrative numerical examples,
we would like to concentrate on the
concept of Bose-Bose dressed Wannier functions,
which appear, e.g., in the {\it self-consistent} treatment of 
mixtures of strongly-interacting Mott-insulators \cite{BB_archive},
and demonstrate how inter-species interaction 
-- in addition to the intra-species interactions --
dresses the bear Wannier functions of each species. 

We consider 1D optical lattices and work in dimensionless units in which 
the one-body Hamiltonian of the $A$-type bosons reads  
$h_A(x) = -\frac{1}{2} \frac{d^2}{dx^2} + 10 \cos^2(x)$
where the depth of the optical lattice is 20 recoil energies.
The same Hamiltonian is taken for the $B$-type bosons.
The intra- and inter-species particle-particle interactions,
$g_AV_A(x-x')$, $g_BV_B(x-x')$ and $g_{AB}V_{AB}(x-x')$,
are taken to be the common contact potential,
see, e.g., \cite{BB_Ho,BB_Esry,BB_Pu}.
The repulsion strengths, $g_{A}$, $g_{B}$ and $g_{AB}$,
are proportional to the corresponding scattering lengths 
and implicitly include the confining parameters of the transverse directions \cite{Maxim_PRL}.
$f_A,f_B$ denote the filling factors.
In 1D one often expresses the strength of the (intra-species) interaction
in terms of the dimensionless parameter $\gamma_{A}$ which is the ratio of interaction and kinetic energies.
In the above units one has $\gamma_A=\pi\frac{g_A}{f_A}$ and similarly for the $B$-type atoms.

We consider the filling factors $f_A=f_B=1$, $N_A=N_B=N_w$ orbitals,
and the strong intra-species interactions $g_A=g_B=10$.
For the inter-species interaction $g_{AB}=1$ 
the dressed Wannier functions of the $A$- and $B$-type bosons are depicted in Fig.~4a.
As can be seen in the figure the two sets of functions sit one atop the other and cannot be distinguished.
For the above parameters and stronger inter-species repulsion $g_{AB}=10$ 
the dressed Wannier functions of each species move apart, see Fig.~4b;
In each site, the two orbitals are pressed one against the other and thus narrow,
and localize at the borders of the lattice site.
Since the inter-species repulsion is quite strong,
the $A$- and $B$-type dressed Wannier functions (orbitals)
would like to reduce their spatial overlap in order to minimize
the interaction in the Bose-Bose multi-orbital energy-functional (\ref{BMF_functional_BB}).
Finally, it is instructive to contrast 
the dressed Wannier functions of the $A$- and $B$-type bosons presented in Fig.~4b 
and the single-species two-band bosons-dressed Wannier functions depicted in Fig.~3.
The latter show negative parts in each well, 
since all $2N_w$ dressed Wannier functions are orthogonal to one another.
In contrast, there can hardly be seen negative parts 
of the dressed Wannier functions of the Bose-Bose mixture,
because dressed Wannier functions belonging to different species need not,
of course, be orthogonal to one another.

\section{Multi-orbital mean-field for Bose-Fermi mixtures} 

The final system we wish to study is a mixture made of 
two kinds of particles of different quantum statistics, 
namely bosons and fermions.
Our starting point is the many-body Hamiltonian describing 
a Bose-Fermi mixture of $N=N_B+N_F$ interacting particles,
$N_B$ spinless identical bosons and $N_F$ spin-half fermions,
in a trap (optical lattice).
We assume for simplicity that the fermionic one- and two-body parts
of the Hamiltonian are spin-independent.
In this case the Hamiltonian is given by,
\beq\label{ham_BF}
\hat H_{BF}(\r_1,\ldots,\r_{N_B},\r'_1,\ldots,\r'_{N_F}) =  \hat H_B + \hat H_F +
 \sum_{i}^{N_B}\sum_{p}^{N_F}\lambda_{BF}V_{BF}(\r_i-\r'_p).
\eeq
Here, both $\hat H_B$ and $\hat H_F$ are the Hamiltonian of single-species interacting particles, 
see (\ref{ham}),
where the corresponding quantities are denoted there by the subscripts $B$ and $F$, respectively.
$\lambda_{BF}V_{BF}(\r_i-\r'_p)$ denotes the inter-species interaction between 
the $i$-th boson and the $p$-th fermion 
where $\lambda_{BF}$ measures the strength of the inter-species interaction.

In the multi-orbital mean-field approach,
similarly to the previous sections,
we attach an {\it orbital} to each of the $N=N_B+N_F$ atoms.
In the simplest case, all bosons reside in the same orbital
whereas each fermion resides, of course, in a different spin orbital.
The corresponding mean-field wavefunction, 
which is symmetric under permutation of any two bosons and
is anti-symmetric to permutation of any two fermions,
is simply given by
\beq\label{GP_HF}
 \Psi(\r_1,\ldots,\r_{N_B},\r'_1,\ldots,\r'_{N_F}) 
 = \phi(\r_1)\phi(\r_2) \cdots \phi(\r_{N_B}) \times 
 \hat{\cal A} \left\{ \psi_1(\r'_1)\psi_2(\r'_2) \cdots \psi_{N_F}(\r'_{N_F}) \right\},
\eeq
where $\hat{\cal A}$ is the anti-symmetrization operator.
This standard mean-field approach has been employed in the 
literature for Bose-Fermi mixtures in traps, see, e.g., \cite{BF1,BF2,BF3}, 
usually when the fermionic atoms are taken to be spin-polarized (also see below).
It may be termed the Gross-Pitaevskii--Hartree-Fock approach for
Bose-Fermi mixtures.

The most general mean-field for the Bose-Fermi mixture is the 
one which allows also for the bosons to occupy different orbitals.
Following the multi-orbital approach in the purely-bosonic case,
we may take $n_1$ bosons to reside in one orbital, $\phi_1(\r)$,
$n_2$ bosons to reside in a second orbital, $\phi_2(\r)$, and so on,
distributing the $N_B$ bosonic atoms among $n_{orb}>1$ orthonormal orbitals.
All together, the multi-orbital mean-field wavefunction for 
the $N=N_B+N_F$ atoms of the Bose-Fermi mixture is the multiplication of 
two single configurations, 
a general permanent for the bosons and a determinant for the fermions,
\beq\label{general_MF_BF}
 \Psi(\r_1,\ldots,\r_{N_B},\r'_1,\ldots,\r'_{N_F}) =
 \hat{\cal S} \left\{ \phi_1(\r_1)\phi_2(\r_2) \cdots \phi_{N_B}(\r_{N_B}) \right\} \times
 \hat{\cal A} \left\{ \psi_1(\r'_1)\psi_2(\r'_2) \cdots \psi_{N_F}(\r'_{N_F}) \right\}.
\eeq 
Of course, the wavefunction (\ref{general_MF_BF}) possesses the appropriate symmetries
with respect to permutation of any two identical particles.
To proceed, we would like to prescribe the energy-functional.
For the bosons we may take any set of $n_{orb}$ orbitals and a corresponding 
set of occupations $\{n_i\}$.
For the fermions, we remind that the Hamiltonian is spin-independent 
and denote in the following quantities related to
spin up and spin down atoms by $\alpha$ and $\beta$ superscripts, respectively. 
The most general single-determinant state is comprised of $N_F^\alpha$ spin up 
fermions and $N_F^\beta=N_F-N_F^\alpha$ spin down fermions,
corresponding to an {\it unrestricted} Hartree-Fock ansatz \cite{Szabo_book}.
It is implicitly assumed that if the fermionic gas is polarized,
say $N_F^\alpha=N$ and $N_F^\beta=0$, then terms involving spin down fermions drop out in what follows. 
With these conventions, 
the expectation value of the 
Hamiltonian with the multi-orbital Bose-Fermi mean-field wavefunction 
(\ref{general_MF_BF}) readily reads:
\beq\label{BMF_functional_BF}
E_{BF} = E_B + E_F + \sum_i^{n_{orb}} 
 \left[ \sum_p^{N_F^\alpha} \lambda_{BF} n_i 
  \int \!\! \int \phi_i^\ast(\r) {\left(\psi_p^\alpha(\r')\right)}^{\!\!\ast} V_{BF}(\r-\r') 
                 \phi_i(\r) \psi^\alpha_p(\r') d\r d\r' + 
\left\{ \alpha \leftrightarrow \beta \right\} \right]. 
\eeq
Here, $E_B$ is the single-bosonic-species multi-orbital energy-functional,
see Eq.~(\ref{BMF_functional}), where the quantities are denoted there with the subscript $B$,
and $E_F$ is the standard, unrestricted Hartree-Fock energy-functional \cite{Szabo_book},
\beqn\label{UHF_functional}
& & E_F = \sum_{p}^{N_F^\alpha} 
\Bigg[ \int {\left(\psi_p^\alpha(\r)\right)}^{\!\!\ast} h_F(\r) \psi^\alpha_p(\r) d\r 
  + \frac{1}{2}\sum_{q}^{N_F^\alpha} \lambda_F  
 \int \!\! \int {\left(\psi_p^\alpha(\r)\right)}^{\!\!\ast}{\left(\psi_q^\alpha(\r')\right)}^{\!\!\ast} 
 V_F(\r-\r') \left\{1-{\mathcal P}_{\r\r'}\right\} \times \nonumber \\
& & \psi^\ast_p(\r) \psi^\ast_q(\r') d\r d\r' \Bigg] + 
  \left\{ \alpha \leftrightarrow \beta \right\} +
 \sum_{p}^{N_F^\alpha}\sum_{q}^{N_F^\beta} \lambda_F
 \int \!\! \int {\left(\psi_p^\alpha(\r)\right)}^{\!\!\ast}{\left(\psi_q^\beta(\r')\right)}^{\!\!\ast} 
 V_F(\r-\r') \psi^\alpha_p(\r) \psi^\beta_q(\r') d\r d\r'. \nonumber \\
\eeqn
The last term in the Bose-Fermi mixture energy-functional (\ref{BMF_functional_BF}) 
represents the interaction between the bosons and fermions.
Finally, it is instructive to remark while examining the unrestricted Hartree-Fock 
energy-functional itself that,
for $\lambda_FV_F(\r-\r')=\lambda_F\delta(\r-\r')$ only interactions between spin up and spin down fermions
contribute to (\ref{UHF_functional}) 
and correspondingly to the Bose-Fermi energy-functional (\ref{BMF_functional_BF}).

The ground-state of the Bose-Fermi mixture is obtained 
by minimizing the energy-functional $E_{BF}$ with respect to
the number $n_{orb}$ of orbitals in which the bosons reside,
their occupations $\{n_i\}$, how many fermions are with spin up and how many are with spin down,
and, of course, the shape of all bosonic and fermionic orbitals.
This results in a set of $n_{orb}+N_F^\alpha+N_F^\beta$ coupled equations that
have to be solved self-consistently,
\beqn\label{BMF_equations_general_BF}
 & & \left\{ h_B(\r) - \lambda_B(n_i+1) J_{B,i}(\r) +
\sum_{j}^{n_{orb}}\lambda_B n_j \left[J_{B,j}(\r) + K_{B,j}(\r)\right] 
 +\lambda_{BF}\left[\sum_{q}^{N_F^\alpha} J^\alpha_{BF,q}(\r) + \right.\right. \nonumber \\ 
 & & \qquad \left.\left. + \sum_{q}^{N_F^\beta} J^\beta_{BF,q}(\r)\right]\right\} \phi_i(\r) =  
  \sum_j^{n_{orb}} \mu_{ij} \phi_j(\r), \qquad \qquad  i=1,\ldots,n_{orb}, \nonumber \\
 & & \left\{ h_F(\r) + \lambda_F 
 \left[ \sum_{q}^{N_F^\alpha} \left( J^\alpha_{F,q}(\r) - K^\alpha_{F,q}(\r)\right) 
       +\sum_{q}^{N_F^\beta} J^\beta_{F,q}(\r) \right] 
+ \sum_{j}^{n_{orb}} \lambda_{BF} n_j J_{FB,j}(\r) \right\} \psi^\alpha_p(\r) = \nonumber \\
 & & \qquad = \sum_{q}^{N_F^\alpha} \varepsilon^\alpha_{pq} \psi^\alpha_q(\r) \longrightarrow 
                                      \varepsilon^\alpha_{p} \psi^\alpha_p(\r), \qquad \qquad 
 p=1,\ldots,N_F^\alpha, \nonumber \\
 & & \left\{ h_F(\r) + \lambda_F 
 \left[ \sum_{q}^{N_F^\beta} \left( J^\beta_{F,q}(\r) - K^\beta_{F,q}(\r)\right) 
       +\sum_{q}^{N_F^\alpha} J^\alpha_{F,q}(\r) \right] 
+ \sum_{j}^{n_{orb}} \lambda_{BF} n_j J_{FB,j}(\r) \right\} \psi^\beta_p(\r) = \nonumber \\
 & & \qquad = \sum_{q}^{N_F^\beta} \varepsilon^\beta_{pq} \psi^\beta_q(\r) \longrightarrow 
                                      \varepsilon^\beta_{p} \psi^\beta_p(\r), \qquad \qquad 
 p=1,\ldots,N_F^\beta. \nonumber \\
\eeqn
Here, $J_{B,j}(\r)$ and $K_{B,j}(\r)$ are the bosonic direct and exchange potentials,
see Eq.~(\ref{BMF_1_body}), where the above $B$ quantities are inserted therein.
Analogously, $J^\alpha_{F,q}(\r)$ and $K^\alpha_{F,q}(\r)$ are defined for the spin up fermions.
The inter-species direct-like potentials are given by,
\beqn\label{BMF_1_body_BF}
& &  J^\alpha_{BF,q}(\r) = \int {\left(\psi^\alpha_q(\r')\right)}^{\!\!\ast} V_{BF}(\r-\r')
 \psi^\alpha_q(\r') d\r', \nonumber \\
& &  J_{FB,j}(\r) = \int \phi_j^\ast(\r') V_{BF}(\r-\r') \phi_j(\r') d\r'. \
\eeqn
For the spin down quantities, interchange $\alpha$ and $\beta$.

Let us examine a few properties of the multi-orbital 
working equations (\ref{BMF_equations_general_BF}) of the Bose-Fermi mixture.
As for the previous cases treating systems made of bosons only, 
the Lagrange multipliers $\mu_{ij}$ are introduced in order to ensure orthonormality
of the bosonic orbitals, $\int \phi^\ast_i(\r)\phi_j(\r)d\r=\delta_{ij}$, 
and satisfy the relations $n_i \mu_{ij}=n_j \mu_{ji}$.
Similarly, the off-diagonal Lagrange multipliers cannot be removed in general.
The Lagrange multipliers $\epsilon^\alpha_{pq}$ and $\epsilon^\beta_{pq}$
are introduced, respectively, to ensure orthonormality of the spin up fermionic orbitals,
$\int {\left(\psi^\alpha_p(\r)\right)}^{\!\!\ast}\psi^\alpha_q(\r)d\r=\delta_{pq}$,
and, independently, of the spin down orbitals, 
$\int {\left(\psi^\beta_p(\r)\right)}^{\!\!\ast}\psi^\beta_q(\r)d\r=\delta_{pq}$.
Each of the sets of fermionic Lagrange multipliers can independently be diagonalized,
because the corresponding direct and exchange potentials are invariant 
under unitary transformations of the orbitals.
This property is indicated by the right-hand arrow in the 
second and third equations of (\ref{BMF_equations_general_BF}),
where the Lagrange multipliers in their diagonal forms are denoted 
by $\epsilon^\alpha_{p}$ and $\epsilon^\beta_{p}$, respectively.
This reminds the situation of the standard, unrestricted Hartree-Fock equations \cite{Szabo_book}.
In solving numerically the multi-orbital Bose-Fermi system (\ref{BMF_equations_general_BF}), 
it is, of course, preferable to work with the diagonal form of the fermionic equations 
[the last two equations in (\ref{BMF_equations_general_BF})] and thereby reduce the computational effort.
In this case, the resulting set of fermionic orbitals are generally delocalized functions.
Having in mind the characterization of (localized) dressed Wannier functions also for the fermions, 
we will keep in the following the Lagrange multipliers matrices
in their full, non-diagonal forms,
$\epsilon^\alpha_{pq}$ and $\epsilon^\beta_{pq}$.

The multi-orbital Bose-Fermi energy-functional (\ref{BMF_functional_BF})
can be employed to study different properties of Bose-Fermi mixtures,
such as quantum phases, the ``pathway to fermionization'' 
(of the bosons in the mixture) and,  
analogously to the case of Bose-Bose mixtures,
demixing of Bose-Fermi mixtures in optical lattices.
Here, we would like to concentrate on optical lattices 
and have a closer look at the dressed Wannier functions.
We consider as an example an optical lattice with commensurate 
filling factors of $f_B$ bosons and one fermion per site.
The fermions are taken to be polarized, say, with spin up, that is $N_F^\alpha=N_F$.
Starting from Eq.~(\ref{BMF_equations_general_BF}) and making use of the 
translational symmetry of the lattice,
we arrive at the final result
for the equations defining the dressed Wannier functions 
of the bosons {\it and} fermions in the mixture,
$\phi(\r)$ and $\psi^\alpha(\r)$, 
which read, $i=1,\ldots,N_w$ and $p=1,\ldots,N_w$:
{\small{
\beqn\label{BMF_dressed_Wannier_BF}
 & & \left\{ h_B(\r) - \lambda_B(f_B+1) J_{B}(\r) +
\sum_{j}^{N_w} \left[\lambda_B f_B \left[J_{B}(\r-\R_j) + K_{B}(\r-\R_j)\right] 
+ \lambda_{BF} J^\alpha_{BF}(\r-\R_j)\right] \right\}\phi(\r)  = \nonumber \\
& & \qquad = \sum_j^{N_w} \mu_{ij}\phi(\r+\R_i-\R_j), \nonumber \\
 & & \left\{ h_F(\r) + \sum_{q}^{N_w} 
\left[ \lambda_F \left[ J^\alpha_F(\r-\R_q) - K^\alpha_F(\r-\R_q) \right] 
+ \lambda_{BF} f_B J_{FB}(\r-\R_q) \right] \right\} \psi^\alpha(\r) = \nonumber \\
 & & \qquad = \sum_{q}^{N_w} \varepsilon^\alpha_{pq} \psi^\alpha(\r+\R_p-\R_q). \
\eeqn
}}
As can be seen, 
the intra-species {\it and} inter-species interactions dress the bosonic Wannier functions 
{\it as well as} the fermionic Wannier functions. 
We remind that for $\lambda_FV_F(\r-\r')=\lambda_F\delta(\r-\r')$ 
the spin-polarized fermions do not interact one with
the other; In this case the fermionic Wannier functions are dressed only due to the
interaction with the bosons.

\section{Concluding remarks}

The multi-orbital mean-field approach 
has been very successful in describing and predicting physical 
phenomena of spinless identical bosons in optical lattices and other traps,
and very recently, also in a first application to bosonic mixtures in optical lattices.
These have motivated us in the present work 
to extend, explicitly derive, and to specifically consider
the multi-orbital mean-field approach for systems commonly studied in the  
cold-atom-physics literature in general and in the optical-lattice community in particular.
Specifically, we have described in this paper by multi-orbital mean-field Ans\"atze
(i) spinless identical bosons, (ii) spinor identical bosons (iii),
Bose-Bose mixtures, and (iv) Bose-Fermi mixtures 
in {\it real space} optical lattices.

In the multi-orbital mean-field approach we attach an {\it orbital} to each {\it particle}. 
What are the 'rules' for constructing 
a multi-orbital mean-field ansatz for a given physical system 
and what is the relation between the different orbitals employed?
First, orbitals belonging to different particles bear no
a priori relation between them and hence are allowed to overlap
(i.e., they need not be orthogonal to one another).
Second, two identical bosons can sit in either the same
orbital or in two orthogonal orbitals.
Third, each fermion, of course, sits in its own spin orbital.
Fourth, for spinor identical bosons (and for fermions), 
it is different two spinors which are orthogonal one to the other;
their individual respective components need not be orthogonal.
Finally, the multi-orbital wavefunction is symmetrized (anti-symmetrized)
to possess the correct permutational symmetry with respect to
interchanging any two identical bosons (fermions).

Having prescribed the multi-orbital wavefunction, the expectation value
of the Hamiltonian is calculated, 
leading to the multi-orbital mean-field energy-functional. 
What are the variational parameters of the energy-functional? 
For bosons, these are the occupation of each orbital (or spinor)
and the number of orbitals (spinors). 
Of course, the orbitals (spinors) of all bosons and fermions are themselves variational parameters which
are to be determined self-consistently.
Minimizing the energy-functional with respect to its arguments leads
to a set of coupled, non-linear, generally integro-differential equations for the orbitals. 
These equations have been explicitly derived for a general
inter-particle interaction for the above-mentioned systems.

Particular attention in the context of optical lattices has been given to solutions 
of the multi-orbital equations which possesses translational symmetry. 
These are the {\it dressed Wannier functions} which appear when interaction sets in.
The dressed Wannier functions are the set of orthogonal, translationally-equivalent  
orbitals which minimizes the energy of the Hamiltonian including 
boson-boson (particle-particle) interactions.
It has generally been shown and specifically demonstrated for spinless bosons and mixtures
how intra-species interactions, inter-species interaction,
and couplings to higher-bands can dramatically alter the shape of the Wannier functions 
in comparison to the bear ones.
Moreover, an interesting picture is expected for spinor particles, 
in which different couplings of the individual spinor components
introduce additional possibilities to dress in a different manner 
each component of the {\it Wannier spinor}. 

All the above findings demonstrate the wide potential of
the multi-orbital approach for cold atoms.
We anticipate that the employment of the
multi-orbital mean-field  Ans\"atze presented here 
is to produce further valuable understanding 
and predictions of the physics of cold atoms in optical lattices, 
as well as in other traps.

\begin{acknowledgments}
\noindent
We thank Markus Oberthaler and J\"org Schmiedmayer for discussions.
Financial support by the Deutsche Forschungsgemeinschaft is gratefully acknowledged.
\end{acknowledgments}


\begin{figure}[ht] 
\vglue  -0.5 truecm
\hglue -1.5 truecm
\includegraphics[width=13cm,angle=-90]{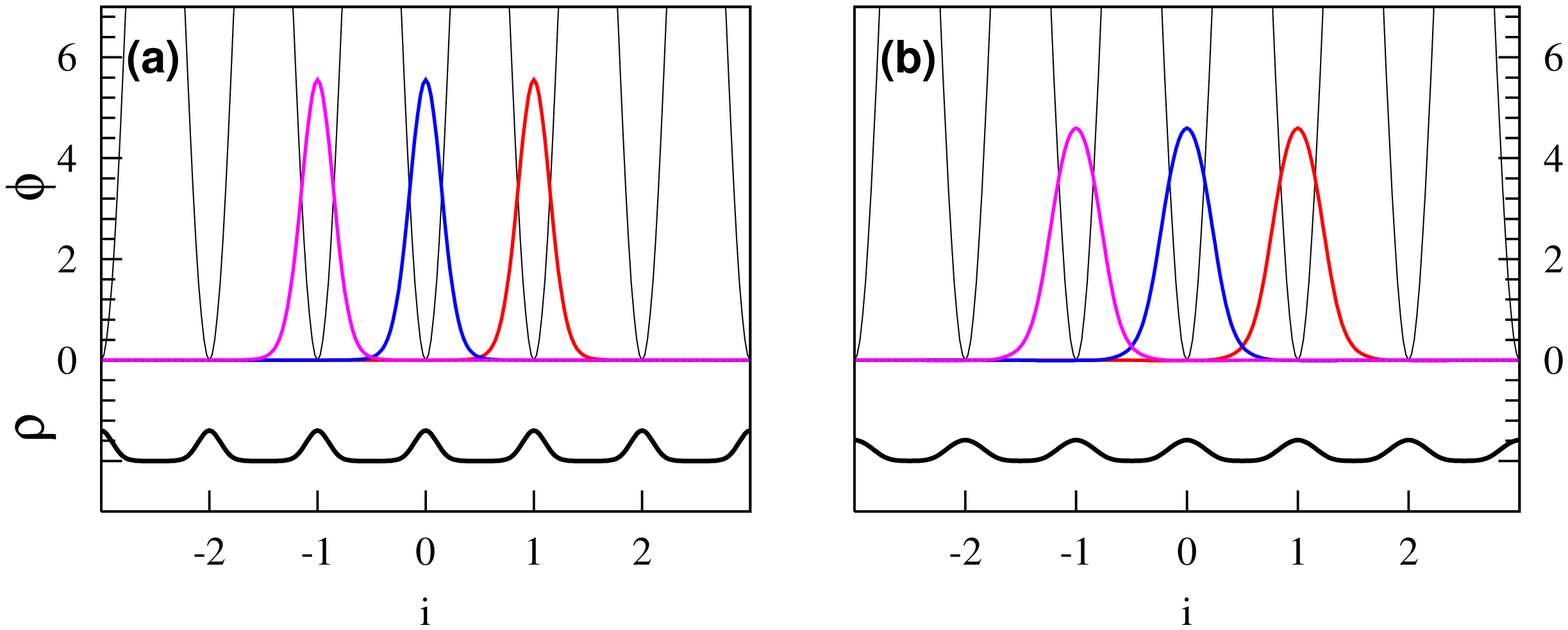}
\caption [kdv]{Boson-dressed Wannier functions in a deep 1D optical lattice. 
Plotted in colors are three adjacent dressed Wannier 
functions $\phi(x)$ versus the site index ``i''.
The optical lattice is illustrated for guidance by the background sinusoidal curve.
Also shown is the spatial density
$\rho(x)=\frac{1}{N_w}\sum_i^{N_w}\phi^2(x-X_i)$, where $X_i$ are the lattice vectors.
For convenience, $\phi(x)$ is normalized on a segment of length $2\pi$.
The optical lattice is of depth $25E_R$.
Other parameters are: (a) Commensurate filling of $f=1$, $N_w=102$ sites and $\gamma=0.00776002$; 
(b) Commensurate filling of $f=2$, $N_w=51$ sites and $\gamma=12.7418$.
The repulsive interaction between two bosons in a site makes 
the dressed Wannier functions wider and increases the spatial overlap
between two neighboring functions. 
See text for more details.
}
\end{figure}


\begin{figure}[ht] 
\includegraphics[width=12cm,angle=-90]{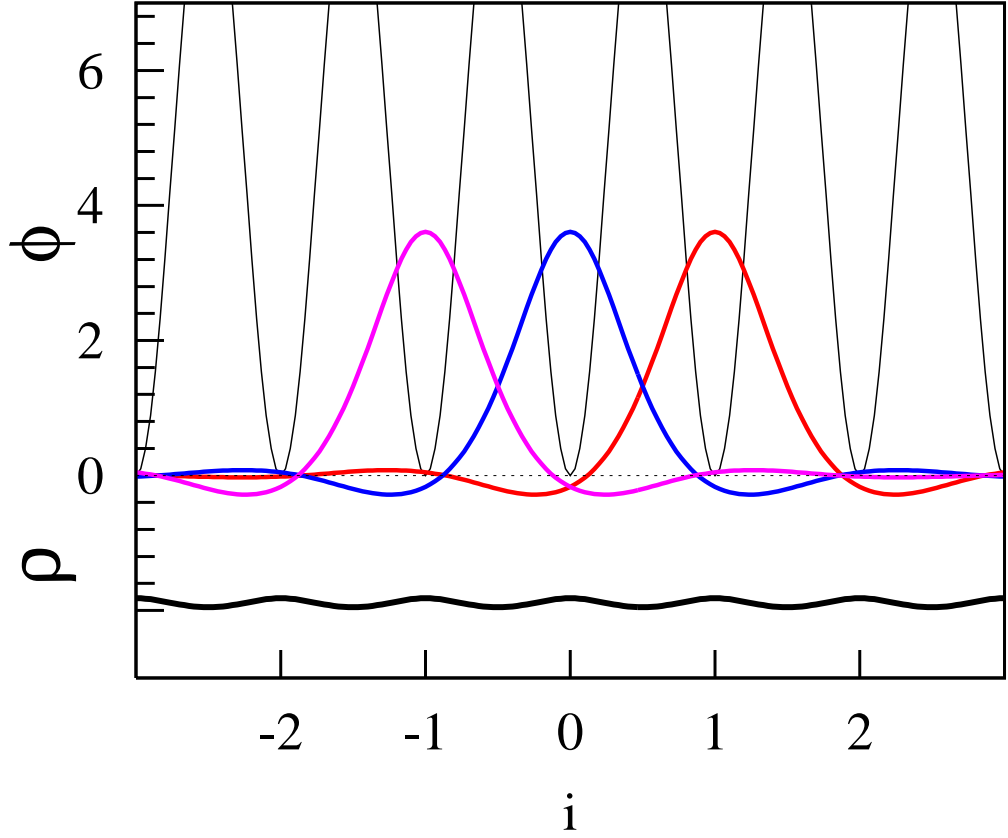}
\vglue  1.0 truecm
\caption [kdv] {Boson-dressed Wannier functions in a shallow 1D optical lattice.
Same as in Fig.~1 except for the parameters: 
Optical-lattice depth of $E_R$, 
commensurate filling of $f=1$, $N_w=102$ sites and $\gamma=3.491$.
See text for more details.
}
\end{figure}


\begin{figure}[ht] 
\includegraphics[width=12cm,angle=-90]{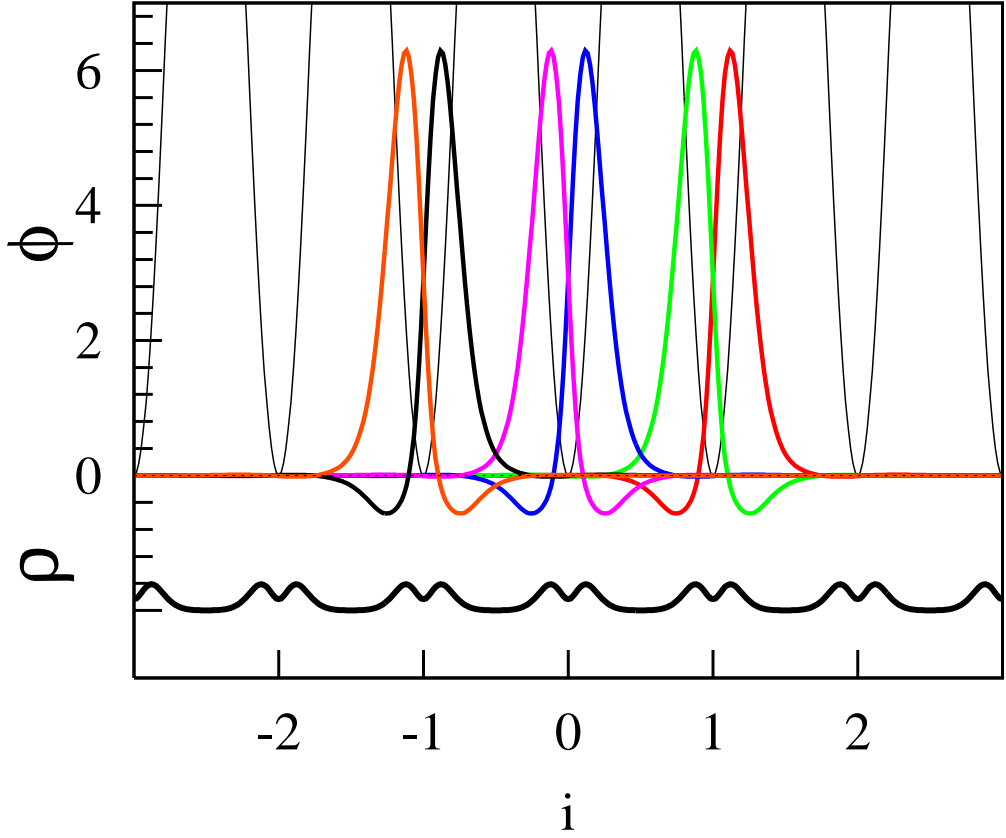}
\vglue  1.0 truecm
\caption [kdv]{Two-band boson-dressed Wannier functions in a deep 1D optical lattice. 
Plotted in colors are six adjacent dressed Wannier functions [three pairs of $\phi(x)$ and $\phi(-x)$] 
versus the site index ``i''.
The optical lattice is illustrated for guidance by the background sinusoidal curve.
Also shown is the spatial density
$\rho(x)=\frac{1}{2N_w}\sum_i^{N_w}\left[\phi^2(x-X_i)+\phi^2(-x-X_i)\right]$, where $X_i$ are the lattice vectors.
For convenience, $\phi(x)$ is normalized on a segment of length $2\pi$.
Parameters are: Optical-lattice depth of $25E_R$, 
commensurate filling of $f=1$, $N_w=51$ sites and $\gamma=12.7609$. 
The two-band {\it dressed} Wannier functions are substantially different
from the {\it bear} Wannier functions of the first and second band.
See text for more details.
}
\end{figure}


\begin{figure}[ht]
\vglue  -2.5 truecm
\hglue -1.5 truecm
\includegraphics[width=13cm,angle=-90]{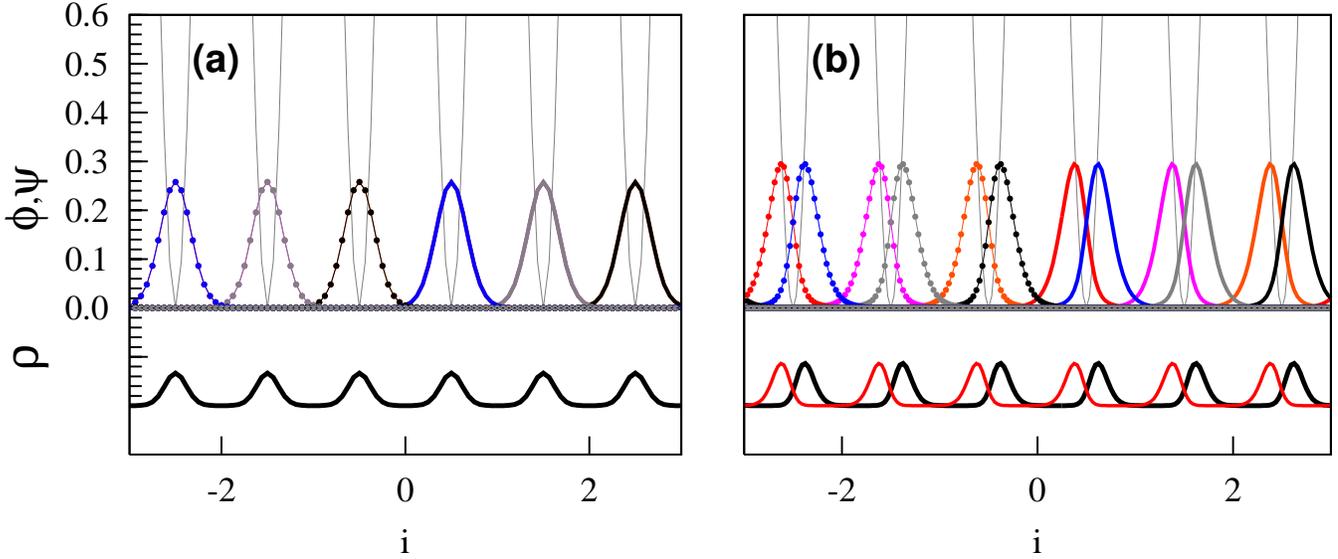}
\caption [kdv] {Dressed Wannier functions of a strongly-interacting bosonic mixture
in a deep 1D optical lattice.
Shown are the dressed Wannier functions $\phi(x)$ (scaled by $\frac{1}{4}$) 
of the $A$-type bosons in black, gray, and blue
and the dressed Wannier functions $\psi(x)$ (scaled by $\frac{1}{4}$) 
of the $B$-type bosons in orange, magenta and red.
The optical lattice is illustrated for guidance by the background sinusoidal curve.
The index ``i'' enumerates lattice maxima.
Also shown are the spatial densities of
the $A$ bosons (in black) $\rho_A(x)=\frac{1}{N_w}\sum_i^{N_w}\phi^2(x-X_i)$
and the $B$ bosons (in red) 
$\rho_B(x)=\frac{1}{N_w}\sum_i^{N_w}\psi^2(x-X_i)$,
where $X_i$ are the lattice vectors.
$\phi(x)$ and $\psi(x)$ are normalized along the lattice.
The parameters are: Commensurate filling factors $f_A=f_B=1$, $N_w=16$ sites,
optical lattice of depth $20E_R$.
The inter-species interaction strengths are:
(a) $g_{AB}=1$. $A$- and $B$-type dressed Wannier functions are indistinguishable.
(b) $g_{AB}=10$. 
The inter-species repulsion is stronger,
causing the $A$ and $B$ dressed Wannier functions to reduce their overlap in space
and consequently become distinct from one another.  
See text for more details.
}
\end{figure}

\ed